**Title Page:**

# Modeling Duelling Contagions of True and False Information in the Face of Inherent Individual biases

**Vaibhav Krishna, PhD***
Human Nature Lab,
Yale University,
Suite 393A, 17 Hillhouse Avenue,
New Haven, CT 06511, USA
vaibhav.krishna@yale.edu

**Prof. Hirokazu Shirado, PhD**
Human-Computer Interaction Institute
School of Computer Science
Carnegie Mellon University, USA

**Prof. Feng Fu, PhD**
Department of Mathematics,
Dartmouth College,
Hanover, NH 03755, USA

**Prof. Nicholas A. Christakis, MD, PhD, MPH**

Human Nature Lab,
Yale University,
Suite 393A, 17 Hillhouse Avenue,
New Haven, CT 06511, USA
nicholas.christakis@yale.edu

**corresponding author*

# Modeling Duelling Contagions of True and False Information in the Face of Inherent Individual biases

**Abstract:**

Advanced digital communication has revolutionized how people create and consume information, making information diffusion an important topic of research for domains from public health to national security. Real-world scenarios of information diffusion often involve competing narratives – true and false – spreading simultaneously. We propose a novel agent-based co-diffusion model, grounded in "*complex-contagion*" and "*spiral of silence*" theories, to capture how network dynamics exploit cognitive biases to shape such interactions. Our findings reveal that manipulative narratives dominate when early spreaders hold them. These network dynamics further exploit inherent cognitive biases to amplify information diffusion regardless of veracity. Further, while favourable previous experience strengthen collective optimism, unfavourable experiences attenuate optimism only modestly. However, we found that early seeding of agents with lower self-censorship not only constrains the spread of manipulation but can also lead to dominance of well-informed populance. This has implications for policies that aim to facilitate healthier discourse, strengthen social cohesion, and ensure equitable access to reliable information.

## 1. INTRODUCTION

An informed populace is essential for a healthy society, yet the rapid spread of misinformation in recent years has raised significant concerns with respect to the integrity of public discourse and collective decision-making[1,2,3]. Social media platforms, such as Facebook, X, and Weibo, have emerged as a prevalent information source for a large part of society, with 62% of the world's adult population consuming news via these platforms as per the current estimates[4]. The low entry barrier for the use of social media has fundamentally changed how information is diffused, where users can disseminate information simply by sharing content. Amplifying these concerns, emerging technologies such as AI content generation are increasingly being used as key tools for shaping public opinion and behavior[5]. These factors, along with characteristics such as limited view format and ideologically segregated social networks, make online social media platforms an attractive target for malicious dissemination of misleading information[6,7], even posing a major threat to national security, political stability[6] [8], and public health[9].

As a result, there is a growing interest in understanding how (mis)information cascades in online social networks, which can be leveraged to contain the spread and impact of misinformation on society[10,11]. Research on the diffusion of misinformation and related concepts, such as rumors, dates back to the 1960s, initially focusing on offline social networks[12]. In recent decades, attention has shifted to online spaces, where misinformation spreads rapidly and at scale. This includes studies tracing the propagation paths of misinformation in online social networks[13], modeling the spread of fake news[14], and comparing the resharing cascades of true versus false content[15]. These insights have been instrumental in informing debunking strategies and designing interventions to curb misinformation before it escalates, minimizing societal harm[11]. Given the limited resources available to governments and platforms, developing a deeper understanding of diffusion processes remains critical. A rich body of work has modeled misinformation spread on social media by drawing analogies to the diffusion of diseases, employing frameworks such as the independent cascade model[16], linear threshold model[17], and epidemiological models (e.g., SIS, SIR[18–20], and Daley-Kendall SIS model[12]).

Yet, most of these approaches have examined misinformation flows in isolation. In practice, online social networks rarely witness information spreading in isolation. Instead, they often host competing streams such as true versus manipulative narratives[15], partisan political campaigns (Democrat vs Republican)[6,21], or health behaviors (pro- vs anti-vaccination)[22,23]. These competing information flows vie for users' limited attention[24], making their interplay central to understanding real diffusion dynamics. While some prior studies have examined the spread of multiple kinds of information, they have largely focused on structural aspects of networks, ignoring the underlying individual-level factors that shape susceptibility and selective sharing.

To address these limitations, we propose a unified *Competing Information – Social Reinforcement and Willingness* (CI-SRW) co-diffusion framework that integrates the network dynamics with cognitive biases, grounded in psychological and behavioral science. The model combines '*complex contagion*'[25], which captures reinforcement through repeated exposure from different sources, with the '*spiral of silence*'[26], which explains how perceived opinion climate influence individuals' willingness to express their beliefs. This unified framework models how competing information interacts with cognitive biases to shape collective belief formation under uncertainty. We further incorporate the effects of prior exposure and social interactions on evolving risk perceptions. The model is validated using data from a large-scale online behavioral experiment with human participants, providing empirical support for the proposed diffusion mechanism. Finally, we conduct counterfactual what-if simulations to examine how the characteristics of initial seed nodes and targeted interventions influence equilibrium outcomes under high-bias conditions. Our contributions are threefold: (1) introducing a psychologically grounded competing-information diffusion model that extends traditional single-contagion approaches; (2) integrating cognitive biases and communication behavior into network diffusion to explain collective opinion dynamics; and (3) demonstrating, through behavioral validation and intervention analyses, how strategic seeding can counter manipulative information and promote accurate collective beliefs.

These results lay a theoretical foundation for analyzing competing information flow and provide a basis for policy designs in situations that require containing harmful information spread through seeding networks with the right information.

## 2. THEORETICAL BACKGROUND

Understanding collective responses to competing information requires uncovering how individual-level processes of social interaction and decision-making give rise to emergent patterns in collective beliefs and societal dynamics [27]. These emergent effects result from the non-linearities in the agent behaviours and social relationships that cannot be inferred directly from individual observations or aggregate statistical models[28]. While controlled behavioral experiments reveal how individuals respond to uncertainty and social influence, they cannot explicitly specify the plausible causal mechanisms. Agent-based modeling (ABM) provides a complementary framework by translating experimentally grounded behavioral mechanisms into computational agents whose local interactions generate population-level diffusion dynamics[29]. This integration enables both empirical validation of the proposed mechanisms and systematic exploration of counterfactual scenarios that are impractical to investigate experimentally, making it an ideal approach for the current study. In the following subsections, we review related work and provide theoretical foundation that defines our proposed model.

### 2.1 (Mis)Information Spread Model

We use the term "misinformation" broadly to refer to any manipulative information propagation - both unintentionally inaccurate content (e.g., rumors or gossip) and deliberate disinformation campaigns designed to mislead and erode trust in institutions. The analogy between the spread of (mis)information in social networks through interpersonal interactions in a fashion similar to a contagion-like process of infectious disease has motivated the application of epidemic models to information diffusion[18]. Early work drew on the Daley-Kendall (DK)[12] and Maki-Thompson (MK)[30] models by directly comparing rumor spreading to epidemic processes. These models addressed some of the limitations of purely structural approaches, such as the Independent Cascade[16] or Linear Threshold[17] models, by incorporating dynamic node behavior rather than a static approach where node state does not change once infected. Subsequent research widely adopted the Susceptible-Infected-Susceptible (SIS) and Susceptible-Infected-Recovered (SIR) frameworks[18], adapting them to model online information diffusion. In these models, the population is stratified into "Ignorants" (similar to susceptible, who have not encountered the information), "Spreaders" (infected individuals who actively disseminate the information), and "Stiflers" (recovered individuals who know but no longer share). Over time, extensions have sought to better capture real-world dynamics. For example, models incorporated incomplete reading behaviors to examine propagation in micro blogs[31], forgetting mechanisms[32], autonomous origins of the spreading phenomenon[33], and time-varying spreading rates to account for user inattention in online envirnments[34].

A further stream of research has extended the epidemic framework by considering the explicit network topologies of affected populations. These studies explore the diffusion process not only by accounting for transmission rates but also by including the structural properties of the underlying graph. Studies have examined how modularity in online communities shapes rumor cascades[35], how overlapping membership across communities influences spread[36], and how multiple routes of transmission affect final rumor size[37]. While these studies highlight the importance of heterogeneity in network structures, they also largely treated individuals as passive carriers of information, where their behavior is determined by network position.

Moreover, the traditional epidemic and network-based models generally focus on the spread of a single piece of information, whereas real-world social media platforms are characterized by simultaneous, often antagonistic, content streams[38]. To address this, recent studies have modeled competing diffusion processes in what have been called "dueling contagions"[39,40]. Wang et al. investigated the parallel diffusion of two misinformations about the same theme[41], while Liu et al. extended the SIR framework by incorporating "hesitator" state to represent undecided individuals for modeling conflicting narratives[42]. Other studies have explored competitive diffusion in product repurchase[43], as well as the co-diffusion of epidemic processes and awareness in multiplex network[44].

While these refinements move closer to real-world information spread dynamics, they largely assume homogeneity of individual behaviour, overlooking the role of cognitive and social factors. To address this, researchers have incorporated mechanisms such as forgetting[32], interaction frequency[34], and trust[45]. For example, models introducing trust between spreaders and ignorants show that trust reduces rumor propagation and delays saturation[45], while others consider educational heterogeneity in determining susceptibility to misinformation[46] or attributes of the concept that is spreading[47]. These contributions underscore that behavioral factors significantly shape diffusion outcomes. However, they typically focus on isolated variables and remain underdeveloped in their integration of established theories from social psychology. Thus, despite advances, current models incompletely capture the interplay between network dynamics and the cognitive biases that mediate real-world information diffusion. Addressing this gap requires a unified framework that embeds competing information processes within networks while grounding agent behavior in empirically supported psychological theories.

### 2.2 The problem of cognitive biases

Misinformation is a widely studied topic in psychology and related fields, providing potential explanations for how and why it spreads on social networks. One stream of research in this area focuses on the appeal of misinformation that triggers emotional responses to cater to the readers' preference for high-arousal news, and which can be responsible for the virality of misinformation online[48]. Another line of studies examines the role of social biases, such as the *bandwagon effect* and *normative social influence theory,* that highlight the tendency of people to conform to ideas of the social group they belong to, irrespective of the veracity of the information[49]. Thus, when individuals are exposed to similar information multiple times through social ties, it grows on their minds through an *illusory truth effect*[50], which can also contribute to the '*complex contagion*' phenomenon. This is further strengthened through other cognitive biases, such as *naïve realism,* whereby individuals believe their perception of reality is more objective and factual, ultimately leading to the *false consensus effect*[51], which further makes them think their beliefs are widely accepted. These biases are not only reflected, but also amplified in the structure of the social network the users belong to.

Finally, these biases work as a reinforcement feedback for individuals' own cognitive biases. For example, while on one hand people tend to notice content they have been thinking about owing to *attentional bias*, on the other hand, they also focus more on the content that aligns well with their prior beliefs due to the inherent *confirmation bias*[52]. Similarly, research on partisanship has highlighted that people firmly rely on news that agrees with them, but rarely seek for disproof of their beliefs, a phenomenon known as *congruence bias*[21]. Studies on crisis situations such as healthcare and the recent COVID-19 pandemic have highlighted that while sound policy would have attempted to minimize mortality by framing guidelines to prevent the worst case, human *optimism bias* led many to act as if the best case was in fact the most likely[53].

Cognitive and affective biases—such as unrealistic optimism[54] and the affect heuristic[55]—lead individuals to underestimate personal vulnerability, particularly during emotionally charged or uncertain situations. These biases function as psychological defenses to mitigate anxiety but can distort judgment. Empirical research suggest that individuals often relying on *positive illusions*[56] to cope with threatning situations, mainting self-perceptions of being more capable, more in control, and less susceptible to harm than others. Research on belief updating further shows that people more readily incorporate favorable information than unfavorable ones, resulting in systematic optimism bias[57,58]. This asymmetry has profound societal consequences: it reduces precautionary actions[59] and, when reinforced through social interactions, fosters collective misperceptions. While optimism can promote resilience and psychological well-being, it can be a deterrent to public health by inducing unhealthy or risky behaviors, as seen during pandemics, like COVID-19[53]. In the early stages of pandemics, studies revealed that the majority of individuals tended to downplay their own risk despite widespread warnings, often influenced by limited information and greater perceived control[60].

### 2.3 The role of personal experience

Though proven to be resistant to change[61], a growing body of reseach supports the notion that personal epxerience moderate the optimistic bias[62]. Experiencing a victimising event often challenges individuals' perceived invulnerability, leading them to recalibrate risk estimates[63]. Early studies among college students found that exposure to a given negative event led students to feel more vulnerable to that event than the average college student[64]. Studies such as the one conducted after the 1989 California earthquake[65] and 1994 Northridge, California earthquake and following five months[62], showed reduced optimistic bias and produced more realistic vulnerability assessments over time. These findings suggest that emotional salience and experiential learning may override the cognitive mechanisms that normally sustain optimism bias. Despite extensive work, most evidence derives from short-lived, localized events (earthquakes or nuclear plant explosions) and individual-level analyses. Less is known about how optimism bias evolves in group contexts, where social reinforcement and information diffusion may magnify or attenuate biased updating. Further, little is known about crises of prolonged and fluctuating intensity, that pose unique conditions under which optimism may resurge or persist despite experience. Addressing these gaps is essential for understanding how collective risk perception adapts over time.

### 2.4 Spiral of Silence

Extant research on mass communication highlights that individuals who perceive themselves as holding a minority viewpoint often silence their opinions. This process is captured by the '*spiral of silence*' theory, which posits that individuals monitor the prevailing opinion climate – driven by fear of social isolation – before deciding whether to express their beliefs[26]. When individuals withhold their views, others underestimate the prevalence of that belief, creating a feedback loop of silence. Empirical studies in online social media extend this insight, revealing the presence of "hardcore minorities", who are disproportionately willing to share unpopular views (e.g., regarding vaccination[66], racism[67], or climate change[68]), while a majority in favour of mainstream or pro-social causes often remain silent. Such dynamics contribute to '*pluralistic ignorance*', where a silent majority wrongly perceives itself to be outnumbered by a more vocal minority[69].

While the spiral of silence has been extensively examined in opinion climate and social media research, it remains largely absent from information diffusion models. Information diffusion studies typically operationalize user silence through mechanical processes such as "forgetting" [32] or "hibernation," where spreaders stop sharing due to memory decay, or through "hesitation" [42], where users pause due to uncertainty about content authenticity. While these extensions acknowledge non-diffusion behavior, they reduce silence to an individual-level stochastic process rather than a socially conditioned decision. Since social influence is central to both *complex contagion* and the *spiral of silence*, their interplay is likely to shape the fundamental dynamics of information diffusion.

Figure 1 illustrates the influence of opinion climate on an individual in the presence of the spiral of silence.

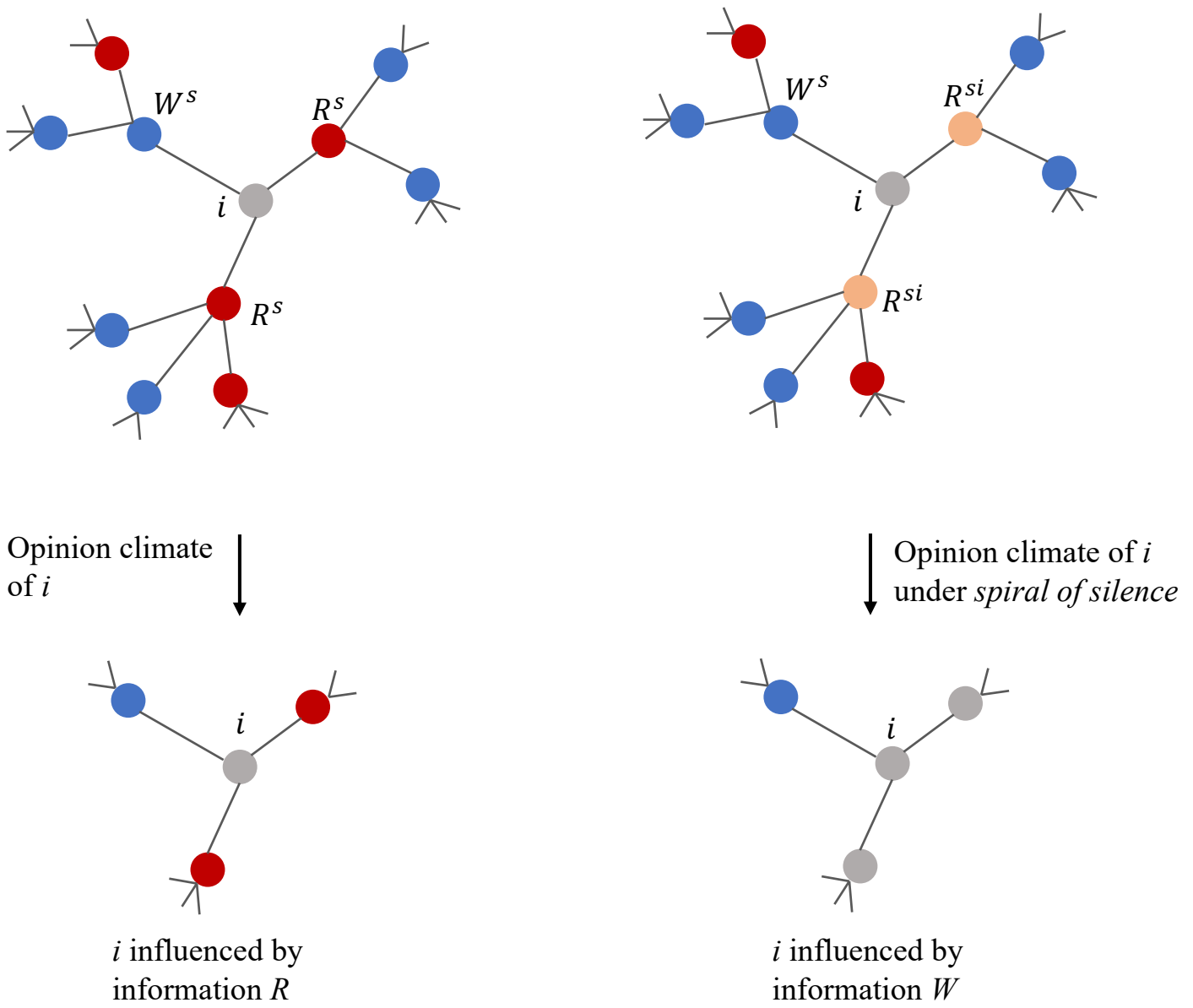

**Figure 1:** Information contagion under spiral of silence

To fill these gaps, we propose a unified co-diffusion model of competing and opposing information, integrating the "*complex-contagion*" and the "*spiral of silence*" theories, to capture how network dynamics exploits cognitive biases in shaping interactions on online social networks. Further, we examine how prior exposure to events and social interactions influence collective belief formation under conditions of uncertainity.

## 3. RESULTS

We evaluated the proposed CI-SRW model through a combination of simulation experiments and behavioral validation with human participants. The simulation experiments examine how competing information evolves under different initial seed distributions and levels of cognitive bias, allowing us to identify the conditions that determine the equilibrium prevalence of true and manipulative information. We then compare the model predictions with data from a single round online behavioral experiments to assess whether the proposed mechanisms reproduce observed patterns of information diffusion. We next compare the model predictions with data from another multi-round online behavior experiments to access whether the proposed mechanism for the role prior experience reproduced the observed patterns. Finally, we use the validated model to investigate counterfactual targeted interventions that influence collective belief formation under competing information.

Across the simulation experiments, we found that both the initial distribution of information and the population's cognitive biases are the primary determinants of diffusion dynamics. Networks initially dominated by manipulative information and characterized by stronger optimism bias consistently converged to equilibrium states with substantially lower levels of accurate information. Under the highest-bias condition, over 90% of the population is susceptible to manipulation for convenient truths (SI Figure S1). In the results that follow, we validate the model through online experiments with real human subjects.

### Single-round online experiment

We analyzed the data from the single-round online behavioral experiment with human participants, designed to simulate decision-making in the face of an unpredictable and sudden "disaster". Figure 2 show the diffusion patterns in a 40-node network as simulated by the model and the observed behavior from the online experiment. (c.f. SI Figure S6 for results from other network conditions 10-nodes, 20-nodes and 60-nodes)

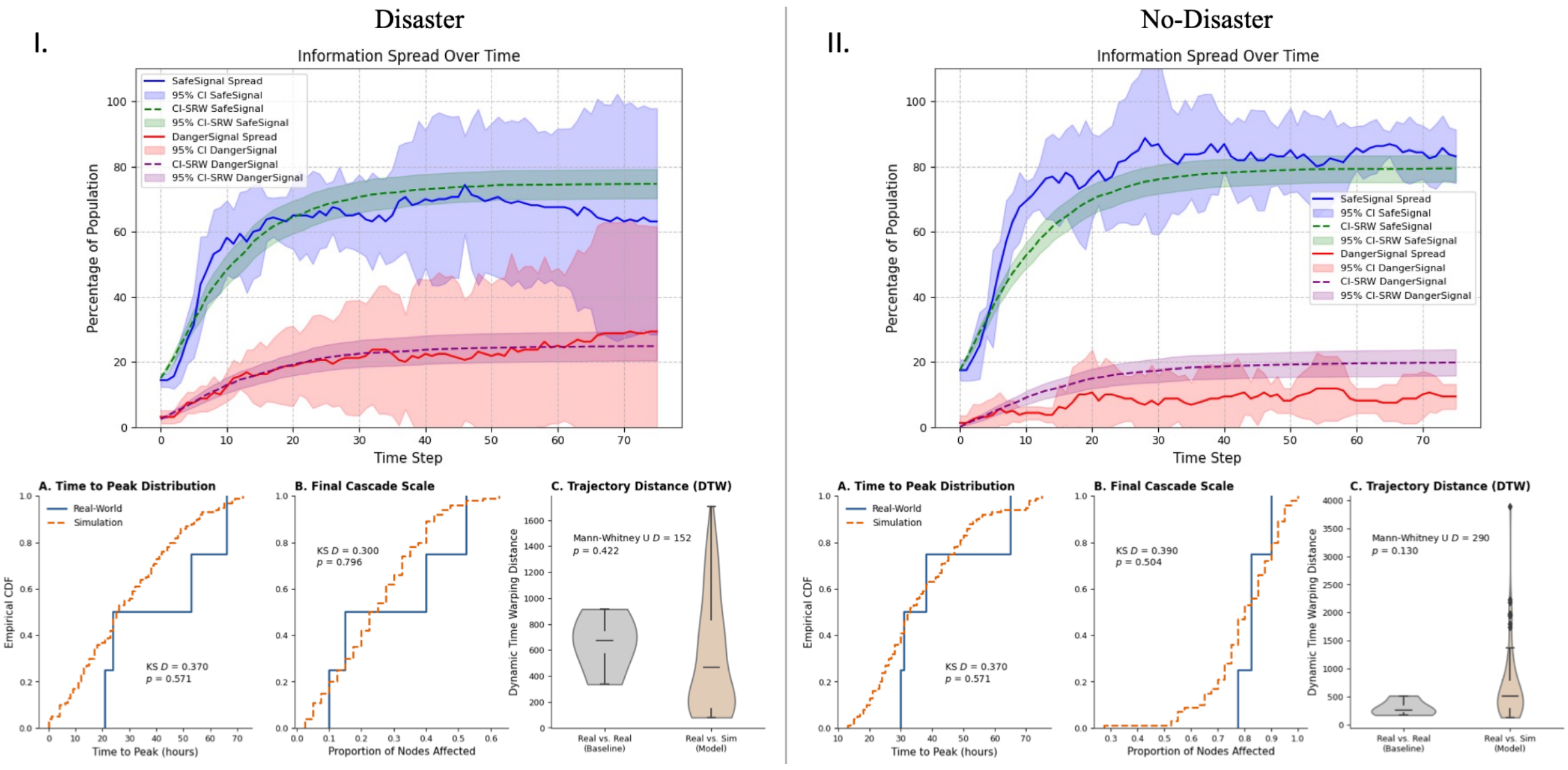

**Figure 2:** Diffusion of 'Safe' (blue and green) and 'Danger' (red and purple) signals with time in a 40-node network. The dotted lines correspond to the simulation result (averaged over 100 simulations) from the proposed CI-SKM model. The solid lines correspond to actual diffusion (average of all sessions in that particular condition) as observed from the online laboratory experiment with real-human subjects. The statistical test subplots (A-C), compares the results from the information flow of the true information – 'danger' signal for disaster scenario and 'safe' signal for no-disaster scenario.

We validated the observed information diffusion pattern at multiple levels. First, we statistically tested the similarity between simulation and experiment outcomes for both the *time to peak* diffusion (Figure 2A) and the *final proportion* of each signal belief (Figure 2B). The comparison for *the time to peak* of diffusion for both 'danger' and 'safe' signals showed no statistical difference in the disaster and no-disaster conditions respectively (*Kolmogorov–Smirnov test*, Figure 2A). Similarly, *the final belief percentage* (proportion of people who believe in 'safe' signal and 'danger' signal) showed no statistically significant differences (with one exception: the diffusion of 'safe' signal in the disaster condition for the smallest network size with 10 nodes; *KS test 0.570, p-value= 0.001*; SI Table S1).

Finally, we compared the full time-series distributions of signal adaptation using the Dynamic Time Warping (DTW) algorithm[70], a widely used and efficient measure of time-series similarity. The comparison of the DTW distance distributions confirmed that the simulated and observed time series were statistically indistinguishable across all the conditions (Figure 2C) (exception for the smaller, 10-node network (*KS-test, p-value=0.003 for safe signal; p-value=0.039 for danger signal*; SI Table S1).

In sum, we find that individual cognitive biases substantially suppress the diffusion of inconvenient truths while simultaneously amplifying the spread of false reassurance, even under crisis. The proposed model reproduces these dynamics in close alignment with the online experiment study. Furthermore, the prevalence of false reassurance is further reinforced as the proportion of correctly informed seed nodes decreases, highlighting the critical role of initial information seeding in shaping equilibrium states of collective belief in the network.

For a robustness check of our findings, we simulated and experimented with individuals in a similar setting, in a regular random-network topology. We find that the simulated and observed time series were statistically indistinguishable across all the conditions (SI Figure S7).

**Role of prior experience - multi-round online experiment**

From the iterated decision-making game, where all subjects played the *evacuation* game lasting four rounds in their assigned group (social network), the collective communication and beliefs in response to uncertain "disaster" was analyzed. The observation patterns were validated at multiple levels. First, we statistically tested the similarity between simulation and experiment outcomes for both the *time to peak* diffusion and the *final proportion* of each signal belief. Finally, we compared the full time-series distributions of signal adaptation using the Dynamic Time Warping (DTW) algorithm[70]. Figure 3 compared the diffusion patterns in scenario 4 (characterized by disaster in round 1 and 3 and no-disaster in round 2 and 4) as simulated by the model and the observed behavior from the online experiment. (c.f. SI Figure S6 for results from other network conditions scenario 1,2,3)

Our findings indicate that individual cognitive biases substantially inhibit the diffusion of unfavorable information, while simultaneously amplifying the spread of reassuring at times potentially misleading narratives, even under crisis conditions. The proposed model successfully reproduces these behavioral patterns, aligning closely with results from the online experiment.

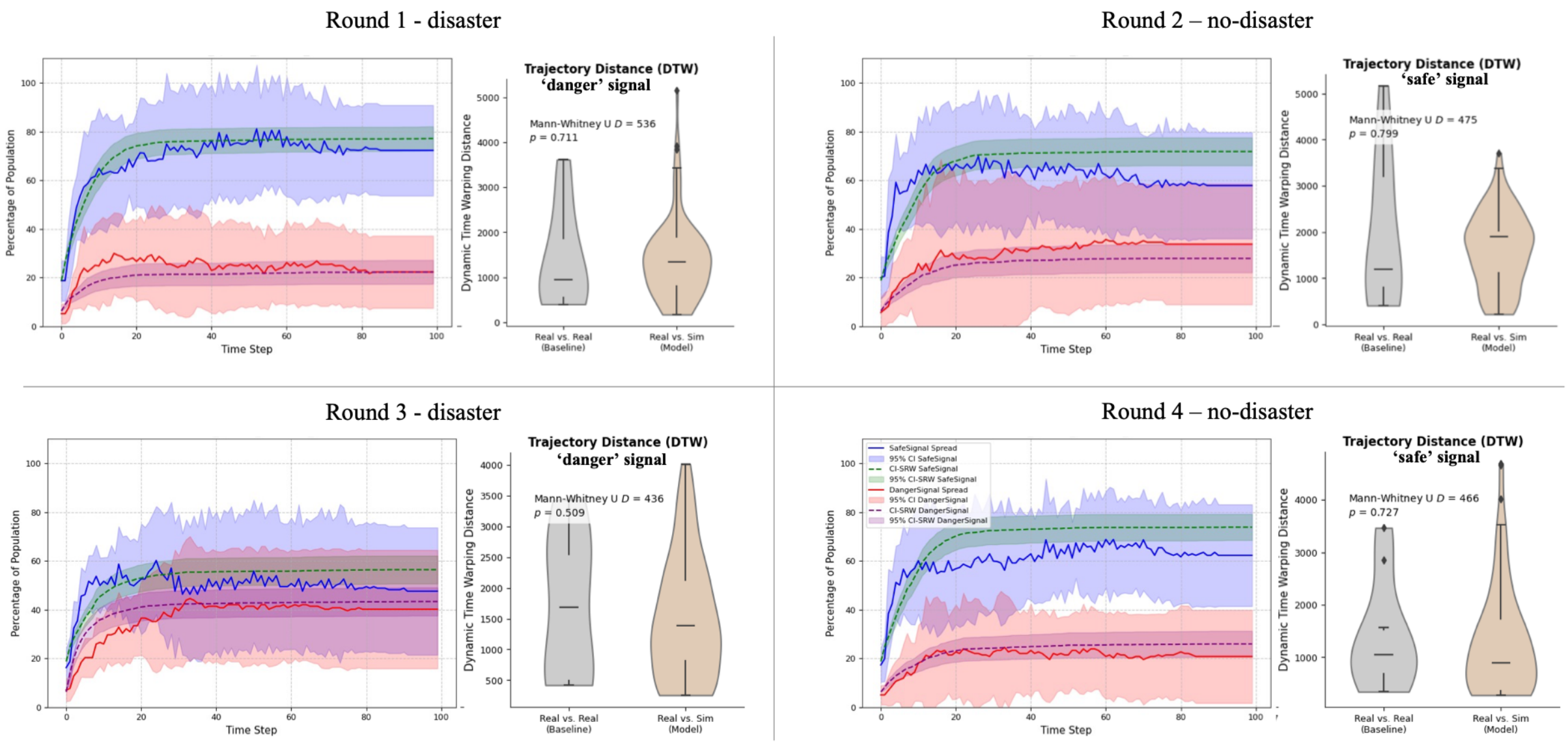

**Figure 3:** Diffusion of 'Safe' (blue and green) and 'Danger' (red and purple) signals with time in scenario-4 characterized by 'disaster' in round 1 and 3 and 'no-disaster' in round 2 and 4. The dotted lines correspond to the simulation result (averaged over 100 simulations) from the proposed CI-SKM model. The solid lines correspond to actual diffusion (average of all sessions in that particular condition) as observed from the online laboratory experiment with real-human subjects. The statistical test subgraphs 'Trajectory Distance (DTW)', compares the results from the information flow of the true information – 'danger' signal for disaster scenario and 'safe' signal for no-disaster scenario.

We further observe that repeated exposure to favorable events strengthens collective optimism about future outcomes, whereas repeated exposure to unfavorable events reduces optimism only marginally at the group level (Scenario 1 vs. Scenario 2) (SI Figure S9a, S9b). Notably, early negative experiences exert a disproportionately strong effect: when such experiences occur before favorable outcomes, they significantly suppress subsequent optimism bias. Consequently, in Scenario 4 (no-disaster rounds following early negative events), a majority of participants fail to maintain "safe" beliefs at equilibrium, whereas in Scenario 3 (no-disaster rounds as early experiences), the proportion of the population with safe beliefs is markedly higher (SI Figure S9c, S9d).

In sum, these results suggest that both temporal sequencing and valence of prior experience play crucial roles in shaping collective belief trajectories, highlighting the persistence of optimism bias even under prolonged exposure to mixed information environments.

### Counterfactual intervention strategies

Using the validated model we simulate two counterfactual intervention strategies to contain spread of misbeliefs during crisis. In the first, we analyse the effect of self-censorship of *warning* informed seeded nodes on steady states. We find that the *willingness to self-censor* of *warning* informed seeds significantly influences the equilibrium state. Figure 4 illustrates the effect of *warning* informed seeds with varying self-censorship on right informed populance in a crisis enviornment.

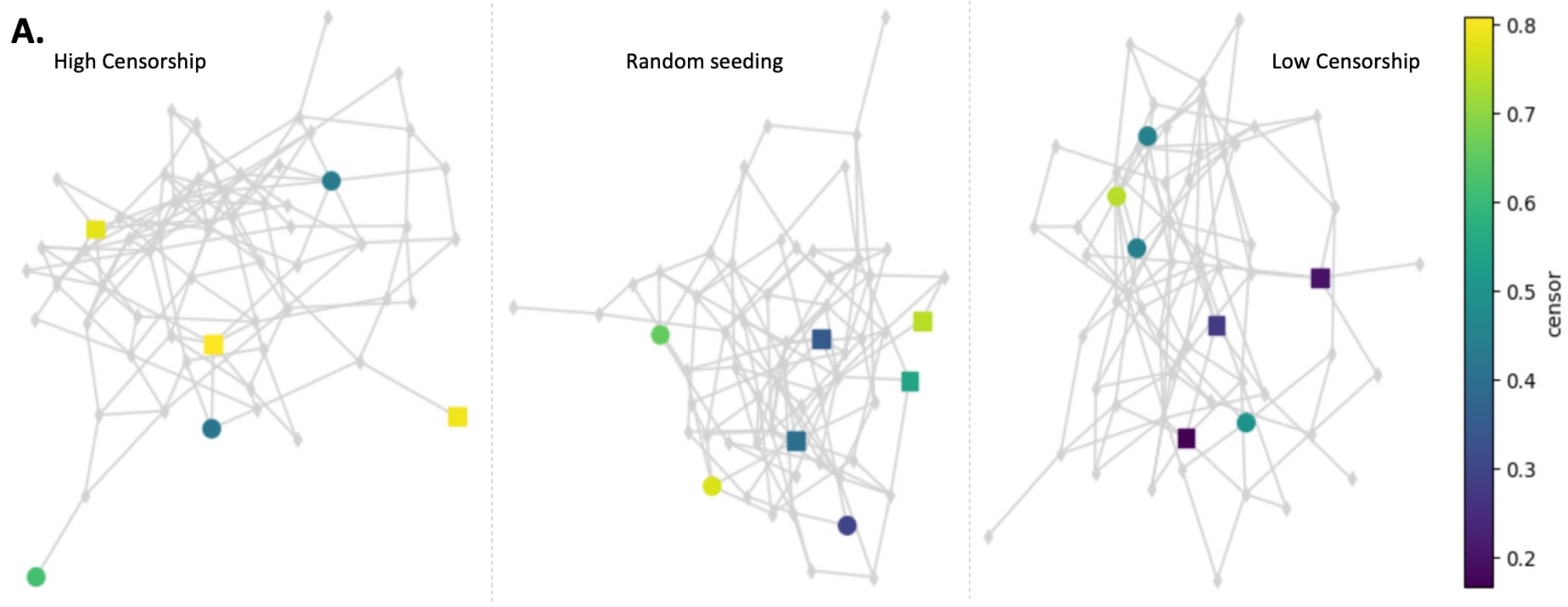

B.
Equilibrium State Comparison: Seeding Efficiency across Densities
Seeding Strategy
Random seeding
High Censorship
Low Censorship
Network Density
⟨k⟩ = 2
⟨k⟩ = 4
⟨k⟩ = 6
⟨k⟩ = 8
⟨k⟩ = 10
Proportion of Right Informed Populance at Steady State (t→∞)

**Figure 4: A.** Illustrates seeding of *warning* information nodes with varying self-censorship in an ER-network. The *reassurance* information is seeded at random. The 'circle' represents seeding nodes with *reassurance* infromation and 'square' represents nodes seeded with *warning* information. **B.** Impact of self-censorship of *warning* informed seed under crisis situation on the right informed populance at equilibrium (*average of 100 simulated networks-1000 nodes ER network*).

Compared to random initial warning-informed seeds, when initial warning-informed nodes show higher self-censorship – mirroring the populance that is relatively shy, more socially anxious, and less argumentative [71] – the proportion of the population holding correct information can decline by up to 50%, under a biased environment. This effect is particularly pronounced in dense networks. In sparse networks, the right informed population decreases significantly- from 43% for networks with random seeds to 31% with high self-censorship seeds (*t-statistics 6.8, p-value < 0.001*). In dense networks, this decline is even sharper, dropping from 38% (random seeds) to 24% (high self-censorship seeds) (*t-* *statistics 8.6, p-value < 0.001*). By contrast, when initially right informed seeds have lower self-censorship – aligning with the individuals having more confidence, higher self-esteem, and thus lower self-censorship – it can alter the equilibrium state, with the majority believing in the correct information. In sparse networks, the right informed populance increases from 43% (random seed) to 51% (low self-censor seeds) (*t-statistics 4.2, p-value < 0.001*). In dense networks, this effect is amplified, with the true-information populance increasing from 38% (random seeds) to 53% (low self-censorship seeds) (*t-statistics 7.2, p-value < 0.001*).

Finally, for the second counterfactual intervention we analyse the effect of adding a fraction of right informed external agents, with no self-censorship, imitating real-world scenarios, where these can be health experts or influential personalities, that can be deploy in a crisis situation to timely intervene in the networks.

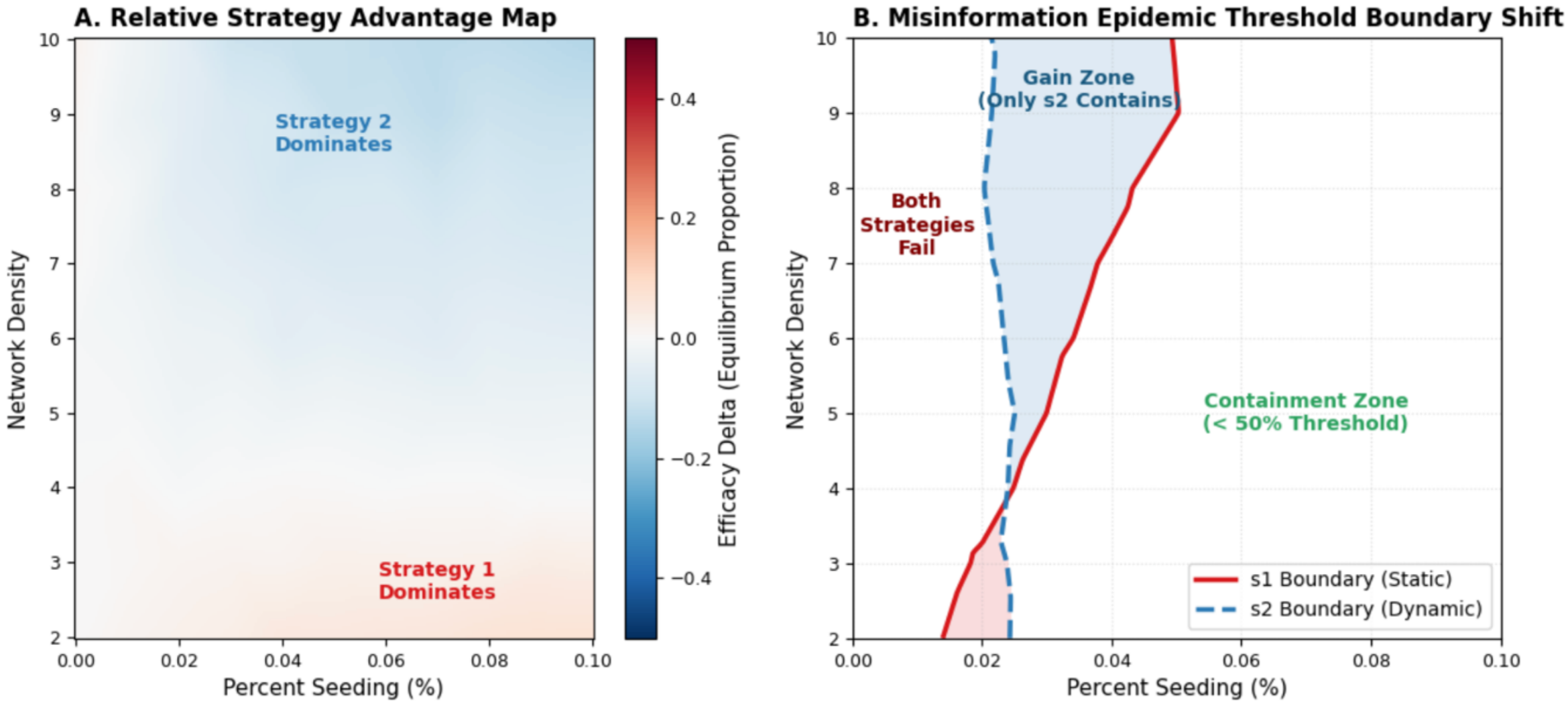

**Figure 5:** Impact of true informed external agents under static (Strategy 1 (s1) – adding four external agents to the network) and dynamic (Strategy 2 (s2) – adding varying number of external agents based on network density) under crisis situation on the right informed populance at equilibrium (*average of 100 simulated networks*). Panel A represent relative advantage of strategy 2 over strategy 1. Efficacy delta is the difference between equilibrium percent of true informed nodes in strategy 2 vs strategy 1. The parameter space in red is where Strategy 1 has advantage, and blue where Strategy 2. In Panel B, the lines are contour level where the steady state proportion of misinformed nodes are 50%. The parameter space on the right/bottom of these contour lines represents misinformed populace falls below 50%. The blue shade region highlights the space where Strategy 1 fails to contain misinformed populace but where Strategy 2 can handle more challenging environment.

We found that, in the fixed-connecting-nodes condition, deploying as few as 2% low-degree external agents (connected with 4 random nodes) can alter the equilibrium state of sparse social networks. However, as the network becomes more dense, the efficacy of the external agents drops, and thus more agents are required to achieve the same equilibrium state. In case of dense networks, 5% of right-informed external agents are needed to increase the right-informed populance to atleast 50% at equilibrium (Figure 5) . In contrast, if the number of nodes each external agent is connected to is proportional to the density of networks, deploying even a small fraction of external agents brings significant improvements in well-informed populance across all networks. This aligns with real-world observations that health experts, public officials, and influential community figures can serve as "anchor nodes" that stabilize discourse and counteract manipulative narratives.

In sum, we find that the silence among the initially truly informed nodes has a pronounced detrimental effect, facilitating the widespread diffusion of manipulative information, which is further amplified in dense networks. However, introducing even a small number of exogenous, well-informed agents provides an effective corrective mechanism. Even when positioned at relatively peripheral locations within the network, these agents are capable of shifting the equilibrium toward accurate information, thereby mitigating the dominance of manipulative narrative.

## 4. DISCUSSION

As the world witnesses the unprecedented growth of online platforms, the ubiquitous spread of information – regardless of its veracity – poses mounting challenges for collective well-being[72]. These challenges underscore the need to better understand the mechanisms of information diffusion, especially in moments of societal crises[53]. In such contexts, online networks often facilitate the simultaneous spread of inconvenient truths (warnings) and misleading reassurances (false safety signals) as a form of "*dueling contagion*", an understudied area. Our study contributes to this space by proposing a novel *Competing Information – Social Reinforcement and Willingness* (CI-SRW) diffusion model that integrates cognitive biases into cascade dynamics. Moving beyond models that treat contagion primarily as a function of structural or content properties, our model emphasizes how deep-seated human tendencies, such as unrealistic optimism, can reshape the trajectory of competing information flows at the collective level.

By explicitly modeling optimism bias, our simulations demonstrate that inconvenient truths are systematically suppressed by false reassurances, especially when individuals are prone to underestimating risks. In such environments, misinformation not only spreads more easily, but it also fills voids left by uncertainty, amplifying rumors that align with optimistic worldviews. Laboratory experiment support these insights, showing that exposure to reassurance messages leads to widespread acceptance of "safe signals," even in scenarios of real disaster. Together, these findings suggest that competing information diffusion is governed not only by network structure but also by the interaction between cognitive biases and social influence.

More broadly, our results extend the literature on collective intelligence by identifying a behavioral mechanism through which social influence can reduce collective accuracy[73,74]. Whereas previous studies have shown that social influence can undermine the wisdom of crowds by reducing the diversity of independent judgments[73], our findings suggest that, under competing information, cognitive biases can further amplify this effect by selectively reinforcing appealing narratives. Consequently, the collective outcome depends not only on the structure of social interactions but also on the distribution of cognitive biases within the population.

A key contribution of our model is the integration of the spiral of silence effects into diffusion processes. Existing cascade models often neglect how willingness or unwillingness to speak up influences collective propagation dynamics. We demonstrate that high self-censorship among right informed initial seeds can significantly weaken the spread of accurate information. When well-informed seeds exhibit higher levels of social anxiety, shyness, or fear of disagreement, the proportion of the population holding correct beliefs can decline by half in biased environments, particularly in dense networks. Conversely, confident, low–low-self-censorship initial seeds can reverse this trajectory, enabling the majority of the population to converge on correct beliefs. This highlights the powerful role of self-censorship in shaping equilibrium states of belief formation.

This study also has limitations. First, while our model captures essential biases, it simplifies the heterogeneity of real-world contexts in which expert credibility, political polarization, and trust in institutions vary dramatically. For instance, interventions by external agents such as health experts may succeed in structurally sparse networks but could fail in polarized ecosystems where institutional distrust undermines message uptake. Second, our simulations assume that agents process information uniformly given their bias estimates, yet empirical work shows that identity, ideology, and cultural narratives modulate susceptibility in more complex ways. Third, our simulations from iterative games, followed similified assumption of increased susceptibility of direct ties of informed agents, which in real world could be moderated by pre-existing trust levels, social and political identity, and the quality of prior institutional response. Forth, our laboratory validation, while encouraging, reflects constrained environments that may not fully reproduce the dynamics of large-scale digital platforms where algorithmic amplification and selective exposure play critical roles.

Finally, like all agent-based models, the CI-SRW framework represents one theoretically grounded generative mechanism capable of reproducing observed collective patterns[27]. Consistent with the multi-realizability principal of real world, alternative behavioral mechanisms may produce similar macro-level diffusion dynamics[75]. Accordingly, the model should be interpreted as one plausible theoretically grounded explanation rather than the unique mechanism underlying competing information diffusion.

Despite these limitations, our work advances the theoretical study of competing information contagion by underscoring the psychological underpinnings of silence, bias, and belief. By unifying insights from cognitive psychology, communication theory, and network science, the CI-SRW model provides a richer account of how "dueling contagions" shape collective outcomes under uncertainty. Practically, the results stress the importance of not only countering misinformation structurally, but also empowering well-informed individuals to voice truths in environments where silence can be more damaging than manipulation. This underscores that combating misinformation requires attention not just to algorithms and policies but also to human communication dynamics, confidence, and bias regulation.

Our findings suggest important design implications for digital platforms. Systems could be engineered to algorithmically elevate credible minority voices, counteracting spiral-of-silence effects and ensuring that truth-bearing but less vocal agents gain visibility. Interface nudges might encourage low-confidence users to share reliable information, while credibility signals could help audiences differentiate reassurance from evidence-based warnings[76]. More broadly, platform policies and design interventions should explicitly account for cognitive biases, reducing the structural advantages that manipulative narratives often enjoy. Embedding such bias-aware mechanisms could strengthen resilience against misinformation and foster healthier online information ecosystems.

## 5. Materials and Methods

We first developed the CI-SRW model, then conducted simulations followed by online behavioral experiments to estimate behavioral parameters and validate the model, and finally we used the validated model to perform counterfactual simulations.

### 5.1 Proposed CI_SRW model

#### 5.1.1 Framework Selection

Uncertain emergency situations – such as epidemics, natural and human-caused disasters, and geopolitical conflicts – pose a great challenge to institutions in ensuring a well-informed populance, which is crucial for the success of welfare policies[39,77,78]. Yet, these uncertain situations are often accompanied by the rapid spread of fake news, fueling misperceptions and jeopardizing societal well-being. Past events, including Hurricane Harvey in 2017 and the H1N1 and COVID-19 pandemics, demonstrated how rumors and false narratives can result in significant loss of life and resources[60]. The underlying human cognitive biases further fuel the diffusion process by shaping attention, adoption, and sharing behavior. Here, we focus on modeling information diffusion in social networks in similar disaster-like situations by capturing the co-diffusion of competing narratives – *reassurance* and *warning* – while explicitly integrating behavioral factors such as social influence and bias.

### 5.1.2 Defining Network

We considered an undirected network $G = \{V, E\}$, where $V$ is a set of $|V| = N$ nodes that each corresponds to an online social network user, and $E \subset V \times V$ is a set of contact relationships between two users, in which two competing types of information (*reassurance*, $\boldsymbol{R}$ or *warning*, $\boldsymbol{W}$) co-diffuse[1]. At any given time, the nodes in the network would either be ignorant or hold the reassurance information, or hold the warning information. The underlying cognitive biases of the users would influence their likelihood of believing either of the pieces of information. For example, previous research on pandemics such as H1N1 and, more recently, COVID-19, has shown that *unrealistic optimism bias* is linked to lower risk perception, influencing the likelihood of believing in anti-vaccine narratives[23,53]. Optimistic errors are reported to be an integral part of human nature, with studies consistently reporting that a majority of the population (as high as 80%) display optimism bias[57]. To account for this phenomenon, we introduce a bias multiplication factor $\boldsymbol{\alpha_j}$ in the spreading model, where *(j-1)* represents the number of prior experiences of similar events, for example, the number of previous epidemic waves.

### 5.1.3 Integrating Psychological factors

A central aspect of information diffusion in social networks is the decision of individuals to share or withhold their opinions. According to the spiral of silence theory, fearing social isolation, some individuals might choose not to express their private opinion if it differs from that of the perceived majority. This willingness to self-censor varies from individual to individual, where some people are generally more likely to express their opinions than others[71]. Following previous work, we operationalise this tendency as a continuous '*willingness to self-censor*' parameter $\Phi_i$ for an individual $i$, that does not change over time[71,79]. This individuals' readiness to voice their opinion in public is often subject to social and political context. In democratic societies, institutional safeguards enable enviornment for people to speak freely without fear of authority retribution[80]. Accordingly, we model self-censorship using a right-skewed beta distribution ($\Phi_i \in [0,1]$: $M = 0.28$; $\text{sd} = 0.16$; $\text{skewness} = 0.6$), producing a population in which most individuals exhibit relatively low levels of self-censorship and a smaller minority exhibits substantially higher reluctance to speak publicly. Alternative distributions can be used to represent more restrictive communication environments characterized by higher overall levels of self-censorship. For example, an authoritative environment which undermines open expression, shifting the distribution toward higher self-censorship across the population[81] (S.I. Figure S4 and Figure S5).

Next, we define the confidence of individuals in expressing their opinion, influenced by the observed majority opinion in their neighbourhoods[82], as proposed by the spiral of silence theory, comparing it to self-censorship. A higher confidence than the self-censorship will increase the likelihood of the individual to express his or her opinion, and vice versa. At any given time, the confidence $c_i(t)$ of an individual $i$ is influenced by the opinion climate $\delta_i(t)$, which is the proportional difference between the perceived opinions of the neighbors

$$\delta_i(t) = \begin{cases} \frac{n_{sim}(i,t) - n_{opp}(i,t)}{n_{sim}(i,t) + n_{opp}(i,t)}, & n_{sim}(i,t) + n_{opp}(i,t) \neq 0 \\ 0, & otherwise \end{cases} \quad (1)$$

where, $n_{sim}(i,t)$ are the number of neighbours of individual $i$ having a similar opinion, while $n_{opp}(i,t)$ are the number of neighbours of individual $i$ with opposing opinion. It is important to note that the ignorant or silent neighbors of individual $i$ do not contribute to the opinion climate.

---

[1] The labels *reassuring* and *warning* refer to the perceived implications of the information rather than its objective truthfulness, allowing the framework to generalize to competing narratives across diverse application domains (reassuring versus warning messages during emergencies, pro- versus anti-vaccination narratives, or competing political viewpoints).

Following this, confidence $c_i(t)$ of individual $i$ at time $t$ is updated as follows: $c_i(t) = (1 - e^{-\hat{c}_i(t)})\,(1 + e^{-\hat{c}_i(t)})^{-1}$, where $\hat{c}_i(t) = max\{\hat{c}_i(t-1) + \delta_i(t); 0\}$. The value $\hat{c}_i(t)$ at time $t$=0 is initialized from a uniform distribution, $\hat{c}_i(0) \sim U(0,1)$. The transformation for $c_i(t)$ ensures that its value stays in the interval [0,1]. As a result, when there are more neighbors expressing a similar opinion as individual $i$ than neighbors who are opposing it, $\delta_i(t)$ will be positive, increasing the confidence $c_i(t)$. Similarly, more neighbors with an opposing opinion than supporting neighbors will result in a negative $\delta_i(t)$, thus decreasing the confidence. Since silent neighbors have no influence on opinion climate $\delta_i(t)$, an individual becoming silenced can trigger a cascade of multiple agents becoming silent or speaking, resulting in the false consensus effect.

Considering the above, at any given time, the individuals in the network belong to one of the states: (i) Susceptible $\boldsymbol{S}$ (*individuals who do not yet believe in any information*), (ii) Reassuring-information informed $\boldsymbol{R}$ (sharing $\boldsymbol{R^s}$, silent $\boldsymbol{R^{si}}$; *individuals who believe and may disseminate reassuring information*), and (iii) Warning-information informed $\boldsymbol{W}$(sharing $\boldsymbol{W^s}$, silent $\boldsymbol{W^{si}}$; *individuals who believe and may disseminate warning information*). Each node is associated with a state:

$$s_i(t) = \left[s_i^S(t),\ s_i^R(t), s_i^W(t)\right] \quad (2)$$

At any given time $t$, informed individuals transition between informed state and silent state $\left[R^s \leftrightarrows R^{si},\ W^s \leftrightarrows W^{si}\right]$ based on the opinion climate of their neighbours. That is, if $c_i(t) > \Phi_i$ the silent individual $i$ will transition to a sharing one and vice versa, such that:

$$s_i^R(t) = s_i^{R^s} + s_i^{R^{si}} \quad (3)$$
$$s_i^W(t) = s_i^{W^s} + s_i^{W^{si}}$$

#### 5.1.4 Competitive Information Spread Model – CI-SRW

Finally, we define the spreading phenomena $[S \rightarrow R, S \rightarrow W]$, grounded in complex contagion, where each individual modifies with some probability its state when exposed to the opinion of its neighbours. Let $p_i(t)$ be the probability mass function of $i$ at time $t$:

$$p_i(t) = \left[p_i^S(t),\ p_i^R(t), p_i^W(t)\right] \tag{4}$$

s.t. $$p_i^S(t) + p_i^R(t) + p_i^W(t) = 1$$

Representing the probability of assuming each of the possible states $(S, R, W)$ at time $t$. The dynamics of the system are given by a random realization for $p_i(t+1)$:

$$s_i(t+1) = MultiRealize[p_i(t+1)] \tag{5}$$

The probability mass function $p_i(t+1)$ is defined by:

$$\begin{aligned} p_i^R(t+1) &= (f_i + \alpha_j.\gamma)\, s_i^S(t) + s_i^R(t) \\ p_i^W(t+1) &= (g_i + \gamma)\, s_i^S(t) + s_i^W(t) \\ p_i^S(t+1) &= (1 - f_i - \alpha_j.\gamma - g_i - \gamma)\, s_i^S(t) \end{aligned} \tag{6}$$

where $\alpha_j$ is the bias-multiplicative factor at experience $j$, $\gamma$ is introduced to account for spontaneously getting infected with information[33], such as in the case of rumor generation, especially if it aligns with ones internal bias and beliefs, and $f_i$, $g_i$ are the spreading functions that account for neighbourhood effect from complex contagion dynamics. Figure 6 illustrates the model and parameter estimates.

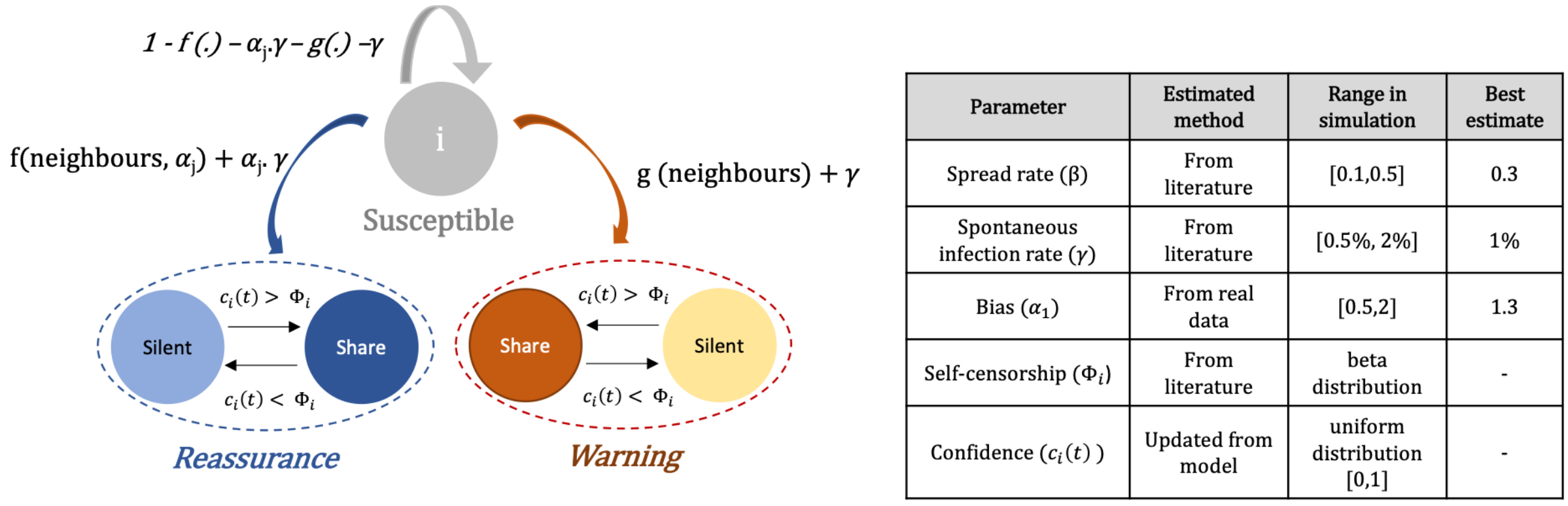

| Parameter | Estimated method | Range in simulation | Best estimate |
|---|---|---|---|
| Spread rate ($\beta$) | From literature | [0.1,0.5] | 0.3 |
| Spontaneous infection rate ($\gamma$) | From literature | [0.5%, 2%] | 1% |
| Bias ($\alpha_1$) | From real data | [0.5,2] | 1.3 |
| Self-censorship ($\Phi_i$) | From literature | beta distribution | - |
| Confidence ($c_i(t)$) | Updated from model | uniform distribution [0,1] | - |

**Figure 6:** The CI-SRW model of competing information flow. The state of an individual can change either by (i) influence from infected neighbors at spread rate $\beta$ and bias $\alpha_j$, or (ii) spontaneously becomes infected at rate $\gamma$ (automatically signal reassurance or warning). The individual begins sharing when the confidence ($c_i(t)$) accumulated through social reinforcement exceeds their inherent self-censorship threshold ($\Phi_i$)

These spreading functions are defined by:

$$\begin{aligned} f_i(t) &= \beta\,\frac{\alpha_j.n_i^{R^S}(t)}{\alpha_j.n_i^{R^S}(t) + n_i^{W^S}(t)} \\ g_i(t) &= \beta\,\frac{n_i^{W^S}(t)}{\alpha_j.n_i^{R^S}(t) + n_i^{W^S}(t)} \end{aligned} \tag{7}$$

where $\beta$ is a constant parameter for the spreading rate, and $n_i^{R^S}, n_i^{W^S}$ are the number of neighbors of $i$ that are informed and sharing it. Note that informed but silent neighbors do not contribute to the contagion process.

### 5.1.5 Role of prior experience in bias update and information spread

The tendency to believe that one's risk is less than that of one's peers is a well-documented and pervasive tendency[57]. However, over the years studies have demostrated lowered optimism bias post crisis experience across health and safety problems[83,84], natural disasters such as tornados[85], earthquakes[62,65], and hurricanes[86], online privacy risks[87] as well as entrepreneurial risks[88]. Several studies report a reduction in optimism bias following direct exposure to adverse events, leading to more realistic future risk estimates. In some cases, the shift may even result in temporary pessimism immediately after the event[89]. Conversely, individuals *only midly* affected by disasters, may experience reinforced optimism, giving a sense of security[90]. Similarly observations were highlighted in studies among healthy women who had more friends with breast cancer, who showed enhanced optimism[91].

Drawing on this evidence, we model bias updating as directionally dependent on prior experience: unfavorable experiences reduce optimism, while favorable ones amplify it. Furthermore, we assume that the magnitude of downward correction following negative experience exceeds the upward adjustment produced by positive experience, consistent with findings of negativity dominance in cognitive processing. We also posit that early exposure exerts greater influence on belief updating than later experiences, given that repeated or prolonged exposure tends to increase familiarity and consequently attenuate perceived risk[92]. Together, these assumptions inform the specification of a bias multiplication factor in our model, capturing how the intensity and valence of prior experience $j$ dynamically shape belief evolution under uncertainty.
We define bias multiplication factor at experience j > 1 as:

$$\alpha_j = \alpha_{j-1} . \exp\left(\left(\frac{(1-2\times d_{j-1})}{j-1}\right) m_d\right) \tag{8}$$

where :
$\alpha_{j-1}$ is the prior bias
$d_{j-1}$ is the indicator for prior experience (1 for disaster and 0 for no-disaster)
$m_d$ is the multiplier for prior experience
Literature on crisis communication suggests mixed insights on how prior experience with public health emergencies shapes individuals respond to expert information. Some studies find that trust in local government is positively associated with perceived health emergency and that prior crisis exposure can strengthen this relationship by reinforcing the legitimacy of official communications[93,94]. Similarly, studies on COVID-19 pandemic showed a net increase in trust in scientists among general public, though this effect was heterogenous, such that individuals with higher baseline trust tent to become more trusting and vice-versa[95]. Overall, the direction of the effect appears contingent on percieved competence of agencies during prior crisis. Given these insights, we adopt a simplifying assumption in our model, such that agents having direct proxy with disaster informed agent, are modelled as having higher initial susceptibility to believing the disaster communication. Thus, we defined information flow at experience j > 1 for individual connected with disaster informed individuals as –

$$g_i(t) = \left(\beta \frac{n_i^{W^s}(t)}{\alpha_j . n_i^{R^s}(t) + n_i^{W^s}(t)}\right) . m_f \quad (9)$$

where :

$m_f$ is the multiplier for disaster informed agent

## 5.2 Experiments

### 5.2.1 Simulation design

First, using computer simulations on both synthetic and real-world networks, we assess the effects of varying initial conditions. We simulated competing information diffusion across two synthetic and one real-world network. For the synthetic network, we selected Erdős-Rényi (ER)[96] random networks and scale-free networks[97], both widely used to approximate structural properties of social, epidemiological, and communication systems. Each synthetic network constitutes 1000 nodes to allow direct comparison. Erdős-Rényi (ER) random networks were generated with an average degree of 4, and scale-free networks were generated where each new node is added to 4 other nodes each time. As a real-world case, we used the Facebook social circles dataset[98], having 4039 nodes and 88324 edges, with an average node degree equal to 43.7. Table 1 summarizes the descriptive statistics of the three networks.

**Table 1: Descriptive statistics of datasets**

| Networks | Nodes | Edges | Average degree | Min degree | Max degree | Average Clustering Coefficient | Modularity Score |
|---|---|---|---|---|---|---|---|
| Scale-free* | 1000 | 3984 | 8 | 4 | 109 | 0.036 | 0.32 |
| Erdős-Rényi* | 1000 | 1998 | 4 | 0 | 12 | 0.004 | 0.52 |
| Facebook | 4039 | 88234 | 43.7 | 1 | 1045 | 0.606 | 0.78 |

**average of 100 simulations*

Within these networks, we seeded 10% of the nodes at random with one of the pieces of information and systematically varied two factors: (i) the initial seed distribution of reassurance versus warning information, and (ii) the population's inherent bias ($\alpha = [1,2]$ - held the same for all nodes in a given simulation).

The spreading function given by the equation-7 model the dynamic contagion behavior, where neighbors' opinions have direct influence on the probability of getting "infected" by a given piece of information. It can be noted here from equation 7 that $f_i(t) + g_i(t) = \beta$, represents the traditional SI model with spreading (sharing) rate $\beta$. Previous studies on information diffusion have highlighted the use of a relatively low spreading rate to isolate the role of individual nodes and structural effect[99]. In our model, the final spreading rate is estimated (equation 7) by incorporating the contagion and bias effects, such that $\beta$ acts as a multiplicative factor incorporating both mechanisms. Following prior literature on information cascades[99 20], we set the value $\beta = 0.3$ in all the experiments. To access the robustness, we varied this value $\beta \in [0.1, 0.5]$. The results indicate that equilibrium outcomes remain largely stable across the range, with the exception of high-bias conditions where very low values of $\beta$ attenuate the diffusion of manipulative information (SI Figure S2).

We also account for the spontaneous emergence of information, a phenomenon commonly observed in social networks where a small fraction of individuals self-generate rumors even in the absence of information[33]. Empirical evidence from extant social media studies shows that less than 1% of users are responsible for more than 75% of low-credibility information on online platforms[100]. To capture this dynamic, we fix the value of parameter $\gamma = 0.01$ in all the experiments (in equation-6).

### 5.2.2 Online experiments

We validated our model against data collected from online experiments – involving 1,240 subjects[101], playing a single round evacuation game and 626 subjects[102], playing multiple rounds, comparing the simulated dynamics with observed behavior.

**Single round evacuation game**

We analyzed data from an online behavioral experiment with human participants, designed to simulate decision-making in the face of an unpredictable and sudden "disaster." The experiment incorporated real economic incentives and networked communication, allowing us to evaluate how interpersonal messages spread over time—an essential process in crisis scenarios. A total of 1,240 participants were recruited via Amazon Mechanical Turk (MTurk) and randomly assigned to conditions within 54 network groups ($N_G = 54$). The reliability of MTurk for behavioral experiments when these data were collected is well established, with prior research showing that participants' behaviors in stylized economic games strongly correlate with those observed in real-world settings[103,104]. To further validate communication effects, a control group of 168 individuals was assigned to an isolated condition where no interpersonal communication was possible. All subjects consented, and the study was approved by the university committee on the use of human subjects.

Following prior empirical work on risk-based decision-making, each participant received a small endowment of US$2.00 at the outset. They then played a time-limited (75 seconds) decision-making game in which they had to choose whether to "evacuate" before an impending "disaster" that might or might not occur. Evacuation required paying a cost of US$1.00. If a participant chose to stay and the game ended without a disaster, they retained their full endowment. By contrast, if a disaster struck while they remained, their endowment was forfeited. The design also introduced an incentive for prosocial information sharing. Whenever a subject avoided a disaster—either by evacuating successfully or when no disaster occurred—they earned an additional US$0.10 for each group member who had taken the correct action. This mechanism represented positive externalities, similar to how individuals in real communities benefit from neighbors' informed decisions, thus motivating participants to disseminate reliable signals about the disaster. Communication during the game was facilitated through simple software buttons embedded in the game interface. Each subject could send messages—"Safe" or "Danger"—to their direct network neighbors, representing their assessment of the likelihood of disaster at that point in time. These signals allowed us to capture the dynamics of information flow under uncertainty, and to examine how local communication shaped collective outcomes.

We conducted 54 network sessions in which 1240 subjects were assigned to different conditions: 24 networks with 10 subjects, 16 networks with 20 subjects, 8 networks with 40 subjects, and 6 networks with 60 subjects; half of the sessions were set up so that a 'disaster' eventually struck, and the other half had no disaster. In each session, subjects were placed in a ring-lattice network where each subject is connected to four neighbors. At the start of the game, one randomly selected subject in each network was told in advance whether a 'disaster' would indeed strike or not, following the real-world scenario where only a small number of people possess accurate knowledge. All subjects were informed that some individuals had accurate information about the disaster and that the immediate neighbors of the informant (always $N = 4$) would know the identity of the true informant. This simulates the real-world situations (such as novel pandemics) where most subjects have no independent way to distinguish fact from opinion. Thus, under such circumstances, individuals resort to social cues and personal preferences in order to make decisions about what to say or do, getting influenced by social and internal biases.

During the session, each individual viewed their immediate neighbors as grey dots in the game window. When any of the neighbors signaled their assessment, by clicking either button, the color of this dot changed for 5s – blue in case of the Safe button and red in case of the Danger button – before automatically returning to grey. Figure 3 illustrates the experimental setup for one of the sessions.
All the subjects passed a series of human verification checks and a screening test of understanding game rules and payoffs. All the subjects consented before the study.

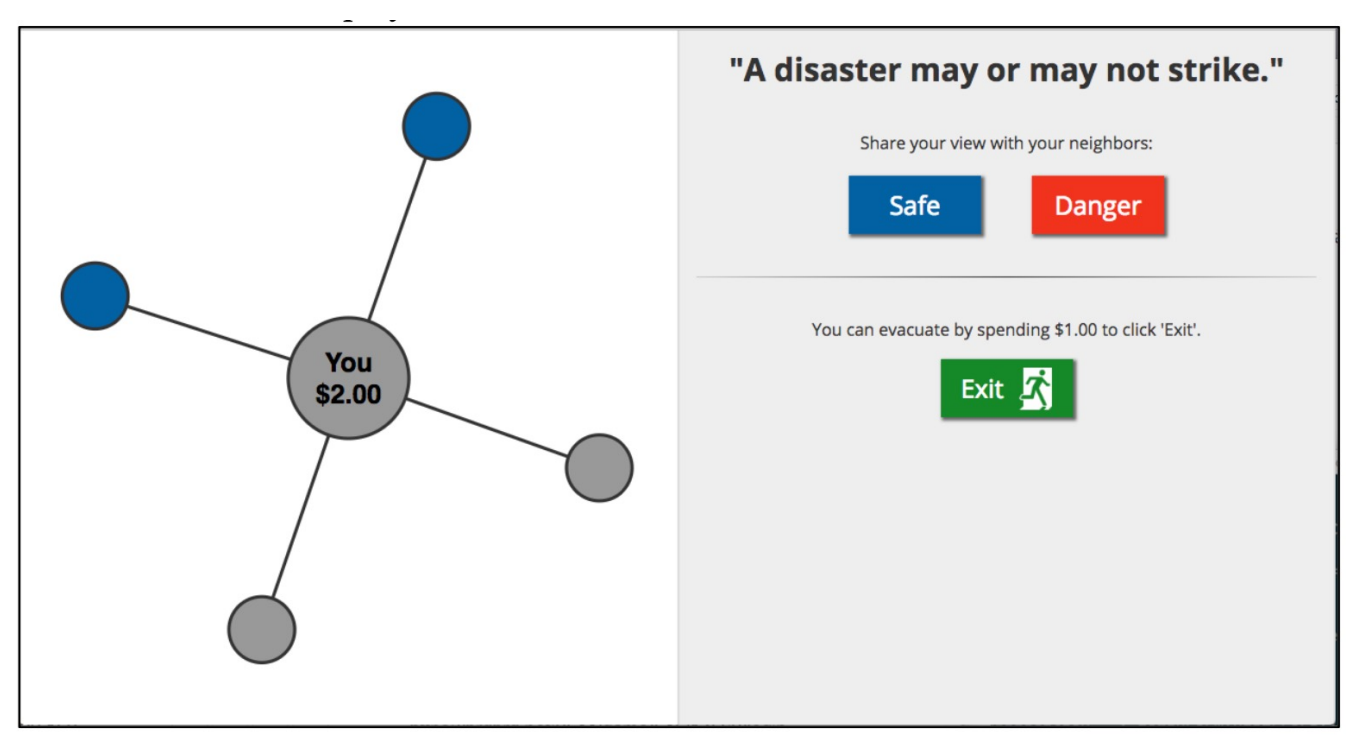

Game view of each player

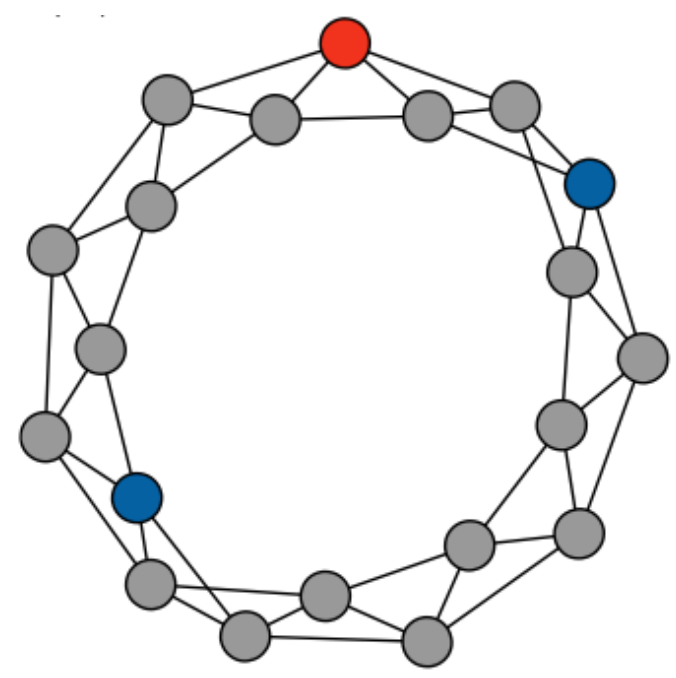

20 node communication network

**Figure 6:** Online experiment setup, with players embedded in a ring lattice of size 20 (subjects do not see the lattice on the right).

We analyzed the networked communication data to validate our competing information diffusion model. Model parameters were estimated from the average behavioral outcomes across sessions under each experimental condition, including the key parameter capturing individuals' underlying cognitive bias. As noted earlier, decision-making in uncertain emergencies is often shaped by optimism bias—a tendency to underestimate risk—which interacts with normalcy bias, leading to false reassurance and delayed protective action[60]. To quantify this bias, we drew on the independent condition, where participants made evacuation decisions without social interaction. Here, the proportion of individuals who chose not to evacuate despite the potential disaster served as a behavioral proxy for optimism bias. Consistent with expectations, higher optimism bias corresponds to a larger share of individuals failing to evacuate. Empirically, we observed that 57.2% of uninformed participants in the independent condition chose not to evacuate. Based on this, we calibrated the bias parameter in our model, setting $\alpha_1 = 1.3$, which we then used in simulations to compare against the experimental outcomes.

To obtain the initial seed distribution, we estimated the average signal count at time $t = 0$ in the sessions for each condition. Table 2 summarizes the estimates of the parameters for different sessions – 10 nodes, 20 nodes, 40 nodes, and 60 nodes under disaster and no-disaster conditions.

**Table 2:** Parameter estimated from the online experiments observations

| *Session* | *Total Nodes* | *Total sessions* | *Percent Initial Seed* (all signal count at t = 0)* | *Percent* of seed nodes sharing 'Safe' signal at t = 0*** |
|---|---|---|---|---|
| **Disaster** | 10 | 12 | 18% | 72% |
| | 20 | 8 | 16% | 76% |
| | 40 | 4 | 19% | 82% |
| | 60 | 3 | 17% | 95% |
| **No-disaster** | 10 | 12 | 18% | 96% |
| | 20 | 8 | 18% | 100% |
| | 40 | 4 | 20% | 94% |
| | 60 | 3 | 17% | 97% |

**values are the average of all the sessions in that particular condition*

We found that, on average, 18% of the nodes acted as initial seeds, transmitting either a 'safe' or 'danger' signal at $t = 0$. Across sessions, the 'safe' signal dominated, ranging from 72% in the 10-node 'disaster' condition to 100% in the 20-node 'no-disaster' condition. Using parameter estimates from the laboratory experiments (Table 2), we simulated each of these conditions to validate the information flow predicted by our model. For comparability, we employed a ring-lattice topology with node degree 4, mirroring the conditions in human-subject experiments. We then compared the temporal dynamics of signal propagation – 'safe' versus 'danger'- generated by the CI-SRW model against the experimental observations.

**Multi-round evacuation game (role of prior experience)**

In the next experiment, we validate the role of previous experience and analyzed the data from a multi-round online behavioral experiment with human participants. An iterated decision-making game was used to study the collective communication and beliefs in response to uncertain "disaster", in a similar *evacuation game.* A total of 626 subjects were placed in 40 groups with an average size of 15.5 (s.d. = 1.1). All subjects played an iterated evacuation game lasting four rounds in their assigned group (social network). The temporal pattern of "disaster" was manipulated, varying its frequency and continuity. A group may experience possible "disasters" across in one of the following arrangements:

**scenario-1**: a "disaster" materialized every round;
**scenario-2**: a "disaster" did not materialize every round;
**scenario-3**: a "disaster" materialized only at round 2 and round 4;
**scenario-4**: a "disaster" materialized only at round 1 and round 3.

In each group, one of the subjects is informed at the start of the game whether a "disaster" would indeed strike. The informed subject could pass the truth about the impeding calamity through network connections. Subject played the game in a directed network with a random graph configuration. All subjects could share the information they believe is true with their neighbors by using "Safe" and "Danger" buttons that indicated their assessment. The bias in each round is estimated using equation-8 and the information flow for disaster informed agent using equation-9. To calibrate the prior-experience multiplier ( $m_d$ ) and the disaster-experience multiplier ($m_f$ ), we performed a grid search over $m_d \in [0.1, 1.0]$ and $m_f \in [1, 3]$. To avoid overfitting, parameter calibration was performed exclusively using data from Scenario 1, Round 2 (disaster event followed by a prior disaster), while all remaining scenarios were reserved for independent model validation. For each parameter combination, we generated 50 independent simulations and compared the resulting distribution of equilibrium true belief proportions with those observed across the 10 experimental groups in Scenario 1 (Round 2) using the Kolmogorov–Smirnov statistic (SI Figure S8). The parameter combination yielding the smallest KS statistic ( $m_d$ : 0.2 and $m_f$ : 1.7) was subsequently fixed and used to simulate all rounds across Scenarios 1–4.

#### 5.2.3 Counterfactual intervention simulations

Finally, we conducted two *what-if* simulations using validated model with Erdős-Rényi (ER) random networks. We generated networks with 1000 nodes, having average degree of $< k >$. To capture the role of network density, we systematically varied connectivity by setting $< k >= \{2,4,6,8,10\}$. For each network, we fixed the proportion of initially informed seed nodes at 10% of the population, with an equal split between those transmitting reassurance information and those spreading warning information. All other model parameters are set similar to the single round online experiment.

For the first counterfactual simulation, we examine the role of self-censorship among the initial *warning informed* seed nodes in shifting the equilibrium between accurate and misleading information. We compared three conditions: (i) higher self-censorship, (ii) lower self-censorship, and a control group. In all three conditions, the seeding nodes for *reassurance information* are drawn at random. The seeding nodes for *warning information* are drawn either at random (control), or nodes with highest self-censorship or nodes with lowest self-censorship. Within this framework, we simulated the diffusion of both *reassurance* and *warning* information, with the former disproportionately favored under stronger optimism bias—for example, accepting false reassurance during imminent danger.

For the second simulation, we evaluate another intervention that examines the effect of adding a fraction of right informed external agents, with no self-censorship. Drawing from the real-world scenarios, these can be health experts, influential personalities, or informed experts that the government or institutions can deploy in a crisis situation to timely intervene in the networks. Within this framework, we examined the effect of adding external agents to the network for two scenarios – (i) each external agent is connected with a fixed small number of nodes in all networks ($< k > = 4$), and (ii) each external agent is connected with $< k >$ existing nodes. Since external agents are added later, their average degree will be lower than the existing users in the network, similar to the real-world scenario, in which, under limited resources, experts can reach to only a small fraction of the population.

Algorithm 1: Competing Information Diffusion algorithm

1. $Initialization$:
    - N agents; with confidence $c_i(t = 0)$ and self-censorship $\Phi_i$
    - Randomly select seed nodes
    - State of nodes:
        - Nodes of susceptible state: S
        - Nodes of reassurance exposed state: R
        - Nodes of warning exposed state: W
    - Bias for R: $\alpha$
    - Spread multiplier: $\beta$
    - Self-generated information probability: $\gamma$

2. $State\ of\ nodes\ after\ time\ interval\ t$:

Define network: $G = \{V, E\}$<br>
Initial state: [S, R, W]<br>
**while** interval<t **do:**<br>
    **for** *i* = 1:N **do**<br>
        observe opinon climate of $i$<br>
        if state = S:<br>
            transfer to R with $\alpha(f_i + \gamma)$<br>
            transfer to W with $(g_i + \gamma)$<br>
            update $c_i(t)$<br>
            if $\Phi_i > c_i(t)$<br>
                stay silent<br>
        if state != S:<br>
            update $c_i(t)$<br>
            if $\Phi_i > c_i(t)$<br>
                stay silent<br>
    **end for**<br>
**end while**<br>
Return state of each node

## Acknowledgments

Funding: We acknowledge support from the Bill and Melinda Gates Foundation. VK acknowledges support from Swiss National Science Foundation (SNSF).

AI Use Declaration: OpenAI was used solely to obtain suggestions for improving the visual presentation of figures without providing any real or confidential data, and to assist with grammar editing.